\documentclass{article}

    \PassOptionsToPackage{numbers, compress}{natbib}

\usepackage[preprint]{neurips_2022}

\usepackage[utf8]{inputenc} 
\usepackage[T1]{fontenc}    
\usepackage{hyperref}       
\usepackage{url}            
\usepackage{booktabs}       
\usepackage{amsfonts}       
\usepackage{nicefrac}       
\usepackage{microtype}      
\usepackage{xcolor}         

\usepackage{enumitem}
\usepackage{amsmath}
\usepackage{amssymb}
\newcommand{\bx}{\mathbf{x}}

\newcommand{\bz}{\mathbf{z}}

\newcommand{\RE}{\ensuremath{\mathbb{R}}}
\usepackage{graphicx}
\usepackage{multirow}
\usepackage{wrapfig}
\usepackage{makecell}
\usepackage{amsthm}
\theoremstyle{definition}

\DeclareMathOperator*{\argmin}{arg\,min}
\usepackage{bbm}
\usepackage{wrapfig}

\title{Evaluating Out-of-Distribution Detectors\\ Through Adversarial Generation of Outliers}

\author{%
  Sangwoong Yoon\textsuperscript{1} \And Jinwon Choi\textsuperscript{2}\And Yonghyeon Lee\textsuperscript{1}\And Yung-Kyun Noh\textsuperscript{3,4} \And Frank Chongwoo Park\textsuperscript{1,5}\\\\
  \textsuperscript{1}Seoul National University, 
  \textsuperscript{2}Kakao Enterprise,
  \textsuperscript{3}Hanyang University, \\
  \textsuperscript{4}Korea Institute for Advanced Study,
  \textsuperscript{5}Saige Research \\
  \texttt{swyoon@robotics.snu.ac.kr}, \texttt{ royce.choi@kakaoenterprise.com},\\
  \texttt{yhlee@robotics.snu.ac.kr}, 
  \texttt{ nohyung@hanyang.ac.kr}, \texttt{ fcp@snu.ac.kr}  
}

\begin{document}

\maketitle

\begin{abstract}
A reliable evaluation method is essential for building a robust out-of-distribution (OOD) detector. Current robustness evaluation protocols for OOD detectors rely on injecting perturbations to outlier data. However, the perturbations are unlikely to occur naturally or not relevant to the content of data, providing a limited assessment of robustness. In this paper, we propose Evaluation-via-Generation for OOD detectors (EvG), a new protocol for investigating the robustness of OOD detectors under more realistic modes of variation in outliers\footnote{The official code release can be found at \href{https://github.com/EvG-OOD/evaluation-via-generation}{https://github.com/EvG-OOD/evaluation-via-generation}.}. EvG utilizes a generative model to synthesize plausible outliers, and employs MCMC sampling to find outliers misclassified as in-distribution with the highest confidence by a detector. 
We perform a comprehensive benchmark comparison of the performance of state-of-the-art OOD detectors using EvG, uncovering previously overlooked weaknesses.
\end{abstract}

\section{Introduction}
\label{sec:intro}

\begin{wrapfigure}{r}{0.5\textwidth}
\centering
  \vskip -10pt
  \includegraphics[width=0.49\textwidth]{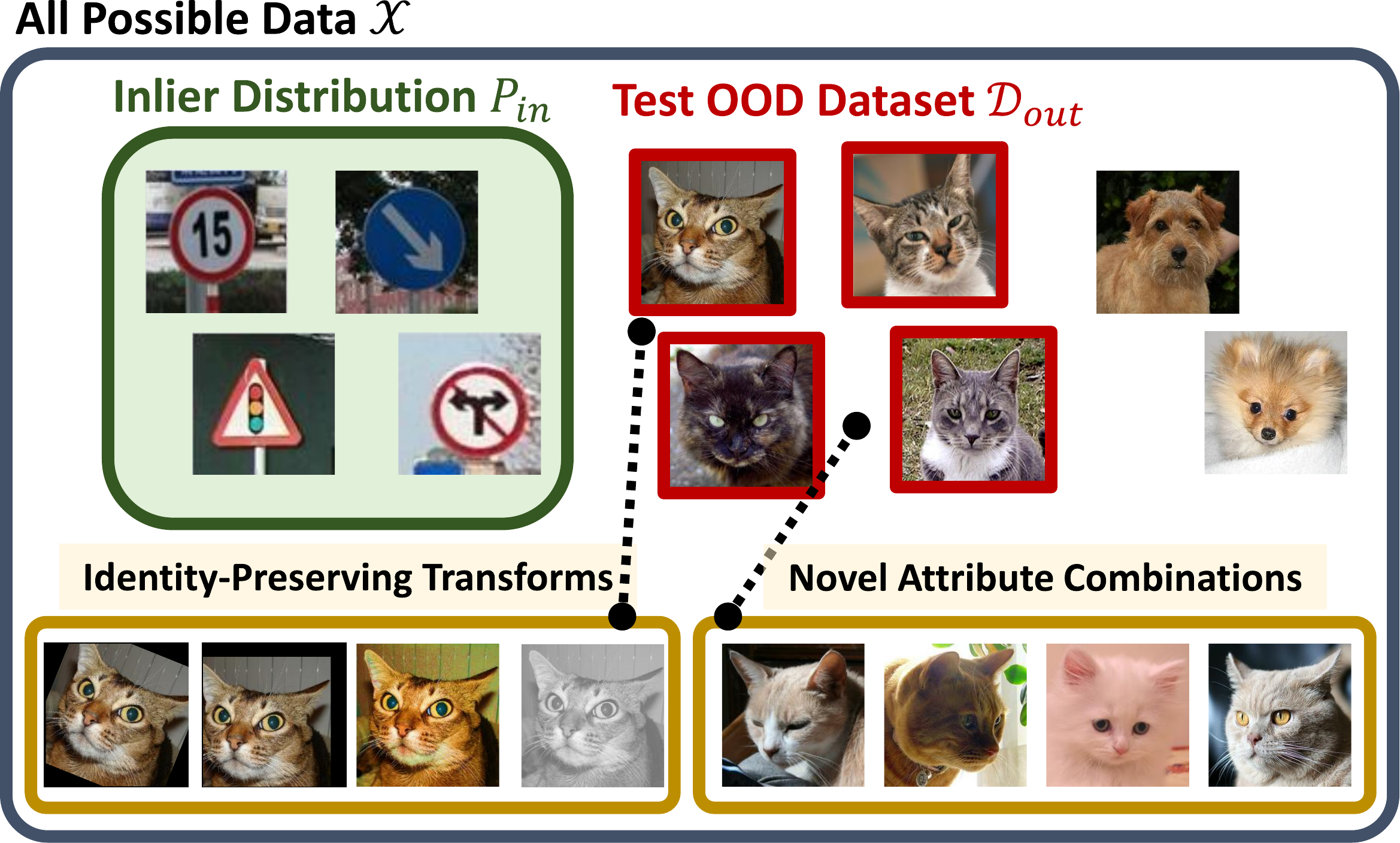}
  \vskip -5pt
  \caption{An illustration of the OOD detection problem setting. The traffic signs are inliers, and cat photos are test OOD data. Cat images in the yellow box are synthesized from the variations of test outliers, residing outside of the test OOD dataset. A robust OOD detector should also detect these images as outliers.
  } 
  \label{fig:illustration}
  \vskip -10pt
\end{wrapfigure}

Outlier detection, also called out-of-distribution (OOD) detection, is concerned with determining whether an input lies outside the training data distribution \cite{hawkins1980identification,hendrycks2016baseline}.
An OOD detector is an essential component of a trustworthy machine learning system, since OOD detection algorithms can pre-filter irrelevant input and prevent the main machine learning model from potentially dangerous erroneous decisions; for example, in a traffic sign classification system, an image of a cat can be filtered by the OOD detector, so that the downstream classifier does not erroneously classify the cat as a traffic sign. 
However, OOD detection is a challenging task, because a detector does not have access to every possible outlier during training.

The diversity of outliers also makes the accurate evaluation of OOD detectors difficult.
In the standard evaluation protocol, an OOD detector is tested using a finite number of test OOD data, for instance a collection of cat photos in the traffic sign classifier case (Figure \ref{fig:illustration}).
An immediate concern for the protocol is that an evaluation result from a test OOD dataset, for example cats, would not always generalize to other outliers, such as dogs.
In practice, this issue has been addressed by using multiple test OOD datasets.

Meanwhile, the non-trivial degree of variability may also present within a single test OOD dataset, creating blind spots in evaluation.
The number of data in a test OOD dataset could be insufficient to cover all possible variations in data, e.g., all possible combinations of eye colors and fur colors in cats, leaving a detector untested for outlier not included in test OOD dataset.
Factors of variations are often continuous in its nature, and multiple such factors form a multi-dimensional space of variations. 
Collecting enough outliers to fill such a space could be expensive.

In this paper, we address this blind spot of evaluation within each test OOD dataset.
We propose \textbf{Evaluation via Generation} for OOD detectors (EvG), a protocol for evaluating the robustness of an OOD detector under a natural variability in a test OOD dataset.
Given a test OOD dataset, EvG first construct a \textbf{variation model}, a set of perceptually relevant variations constructed from the test OOD dataset.
EvG then assesses the robustness of an OOD detector by searching for its worst-case failure mode within the variation model, i.e., an outlier that is classified as in-distribution by the detector with the highest confidence.
The search process in EvG is essentially a black-box optimization, where in each iteration EvG generates a new outlier from the variation model and use the outlier to test the OOD detector.

A variation model is constructed using a generator function $g$, a mapping from a low-dimensional continuous vector space to the data space, where the choice of a generator produces different variation models.
In our experiments for image data, we demonstrate two options for a generator. One is based on an image transformation that preserves the content of an image invariant, and the other is a StyleGAN2 model \cite{Karras_2020_CVPR}.
Both choices of a generator give interesting variation model where we can evaluate the robustness of OOD detectors.

Finding the worst-case failure mode in a given variation model is not trivial because of multiple local optima.
Therefore, instead of
using standard optimization methods that may get stuck at local minima,
we leverage a global search algorithm based on a probabilistic
formulation similar to simulated annealing 
\cite{kirkpatrick1983optimization,metropolis1953equation}.
We first devise an \textbf{adversarial distribution}, a distribution designed to assign a higher probability to outlier samples in a variation model that are likely to be misclassified by the detector.
Then, Markov Chain Monte Carlo (MCMC) is applied to traverse distant modes in the distribution.

EvG can be viewed as an adversarial attack against OOD detectors. However, EvG extends the conventional adversarial attack protocol in two aspects.
First, our variation models cover perceptually meaningful manifold of data, unlike noise-like perturbations used in previous attacks \cite{Hein_2019_CVPR,bitterwolf2020certifiably,sehwag2019rowl,chen2021atom}. Also, our variation models have significantly lower dimensionality than the previous threat models used for OOD detectors, while demonstrating a similar degree of effectiveness in attack.  
Second, optimization is re-formulated as a sampling which is more suitable for global optimization.
Optimization and sampling have an interesting connection, and sampling can sometime be more effective in optimization \cite{ma2019sampling}.

Using EvG framework, we examine the robustness and the failure modes of more than 13 existing OOD detectors.  Among the tested detectors, at least seven detectors
report near-perfect OOD detection performance on a popular benchmark,
CIFAR-10 (in) vs SVHN (out), effectively being indistinguishable with
respect to their performance.  Our investigation reveals that the
detectors in fact have diverging degrees of robustness
outside the SVHN test set.

Our main contributions can be summarized as follows:
\begin{itemize}[leftmargin=1em,topsep=0pt,noitemsep]
\item We propose a novel generative framework to evaluate the robustness of OOD detectors under more natural variations of test OOD data;
\item We provide extensive benchmark results for the state-of-the-art OOD detectors using EvG;
\item EvG reveals previously overlooked failure modes in detectors, disclosing a new room for improvement in OOD detection.
\end{itemize}

\section{Related Work}

\textbf{OOD detection.} The attention on OOD detection has been increasing with an increasing needs for the safety and the robustness in machine learning. Table \ref{tb:ood-detectors} provides a non-comprehensive list of OOD detectors and more discussions on other detectors can be found in Appendix. Most successful approaches in OOD detection are using auxiliary OOD data to calibrate confidence \cite{hendrycks2016baseline,Hein_2019_CVPR,bitterwolf2020certifiably,sehwag2019rowl,chen2021atom}, using Mahalanobis distance in the feature space of pre-trained network \cite{lee2018simple,fort2021exploring}, deep ensemble \cite{lakshminarayanan2017simple}, and self-supervised learning \cite{tack2020csi,sehwag2021ssd,khalid2022rodd}.
OOD detectors based on these ideas all achieves near-perfect (>0.99) AUC score in CIFAR-10 vs SVHN benchmark. One of the key motivations of this research is to investigate the difference in the robustness among these detectors.

\textbf{Robustness of OOD Detectors.}
As other deep neural networks, OOD detectors are not robust to the adversarial attack unless adversarially trained \cite{Hein_2019_CVPR}.
Various techniques have been proposed to improve the robustness
of OOD detectors against small adversarial perturbations, including
adversarial training \cite{Hein_2019_CVPR,chen2021atom}, modification of
the last layer \cite{Meinke2020Towards}, interval bound propagation
\cite{bitterwolf2020certifiably}, and an outlier mining method with adversarial training
\cite{chen2021atom}. 
However, most of such methods evaluate the robustness of a detector only on small perturbations in a norm ball.
In a few notable exceptions \cite{chen2021atom,khalid2022rodd}, the adversarial corruption noise \cite{hendrycks2019benchmarking} which is controlled by discrete parameters.
In this work, we aim to evaluate the robustness of OOD detectors on more natural perturbations of data which are controlled by continuous parameters.

\textbf{Threat Models Beyond Norm Ball.} More diverse threat models have been investigated in the adversarial robustness community, but such techniques are not yet applied to evaluate the robustness of OOD detectors. GAN is employed to generate adversarial examples beyond the small norm ball threat model \cite{song2018constructing}, but it requires an additional classifier and the quality of the generated samples is poor. However, the paper does show that a model robust to noise-like attack is not necessarily robust to other types of attack. The ideas of manipulating an image in a semantic code space to fool a classifier \cite{joshi2019semantic} and parameterizing an attack in the image transformation space \cite{chen2020explore,yang2021random} are investigated in supervised learning setting but for OOD detectors.

\section{Preliminaries: Out-of-Distribution Detection}

\label{sec:background}

\subsection{Problem Setting}

We represent a datum as a $D$-dimensional real-valued vector $\bx \in \mathcal{X} \subset \RE^D$, where $\mathcal{X}$ is the set of all possible values of a valid datum. For example, when $\bx$ is a $32\times32$ RGB image, $D=32\times 32\times 3=3{,}072$ and $\mathcal{X}=[0, 1]^D$. $\mathcal{X}$ is assumed to be compact.

We are given a set of data which are considered as inlier or in-distribution $\mathcal{D}=\{\bx_i\}$.
The samples $\bx_i$ are assumed to be generated from the inlier probability distribution $P_{in}$.
The support of $P_{in}$ is $\mathcal{S}_{in}=\{\bx|\; p_{in}(\bx)>0\}\subset \mathcal{X}\subset\RE^D $, where $p_{in}$ is the inlier probability density.
A vector $\bx$ is an outlier, or an \textbf{out-of-distribution (OOD)} sample, if $\bx$ does not belong to $\mathcal{S}_{in}$ \cite{scholkopf2001estimating}. In plain words, we define an outlier as a sample that has no possibility of being generated from the inlier probability distribution.

An \textbf{OOD detector} $f:\RE^D\to \RE$ is a function which outputs a larger value for an input more likely to be an outlier. 
A test sample $\bx^*$ is classified as OOD if $f(\bx^*)>\eta_f$ for some threshold $\eta_f$.
The function value $f(\bx)$ will be referred to as a \textbf{detector score}.
We shall assume $f(\bx)$ is bounded. 
In this paper, we consider a \textit{black-box} setting, where any information other than the function value $f(\bx)$, such as its gradient, is not accessible.

\subsection{Evaluation of OOD Detectors} \label{ssec:eval}

The currently accepted evaluation protocol for OOD detectors relies on one or multiple \textbf{test OOD datasets} $\mathcal{D}_{out}$, considered as OOD by human prior knowledge.
How well a detector $f$ binary-classifies between $\mathcal{D}_{in}$ and $\mathcal{D}_{out}$ is considered as an indicator of the OOD detection performance.
The classification result is summarized using metrics such as the area under the receiver operating characteristic curve (AUROC or AUC).

Similarly to the adversarial robustness of supervised classifiers (e.g., \cite{szegedy2014intriguing,croce2020robustbench}), the robustness of OOD detectors are evaluated using the worst-case samples within a set of outliers $\mathcal{T}$ defined from the test OOD datasets.
A common choice of $\mathcal{T}$ in previous works on robust OOD detection is a $l_\infty$-norm ball around an OOD datum $\bx_{out}$, i.e., $\mathcal{T}_{\infty}(\bx_{out})=\{\bx |\; \Vert \bx - \bx_{out} \Vert_{\infty} \leq \epsilon \}$, where $\epsilon>0$ is a small constant. 
Then the most confusing outlier to $f$ is found via:
\begin{align}
    \bx_{adv} = \argmin_{\bx} f(\bx) \quad \text{such that} \quad \bx \in \mathcal{T}_{\infty} (\bx_{out}). \label{eq:advattack}
\end{align}
The classification performance on $\bx_{adv}$ serves as a measure for the robustness \cite{bitterwolf2020certifiably}.

Even though an OOD detector is assured to be robust against small-norm perturbations, it is difficult to conclude that the detector is also robust to other types of outliers. 
In particular, the set $\mathcal{T}_{\infty}(\bx_{out})$ is too small to cover $\mathcal{S}_{out}$.  
In the subsequent section, we define a larger set $\mathcal{T}$, called a variation model, by using natural variations of data.

\begin{table*}[t]
\caption{The list of OOD detectors used in the experiments. NLL: Negative Log-Likelihood. RE: Reconstruction Error. MSP: Maximum Softmax Probability. Inference time is the time required to perform inference for 10,000 CIFAR-10 test samples on Tesla V100.}
\label{tb:ood-detectors}
\begin{center}
\begin{footnotesize}
\setlength\tabcolsep{3pt}
\renewcommand{\arraystretch}{1.2}
\begin{tabular}{llcccr}
\toprule

OOD Detector & Description & $f(\bx)$ & \makecell{Inference \\ Time (sec)} & \makecell{Additional\\ Data} \\

\midrule
Glow \cite{kingma2018glow} & A flow-based generative model. & NLL & 93.7 & $\times$ \\
PXCNN \cite{salimans2017pixelcnn++} & PixelCNN++. An autoregressive generative model. & NLL & 32.9 & $\times$ \\
AE (\cite{rumelhart1986}) & An unregularized convolutional autoencoder. & RE & 3.5 & $\times$  \\
\midrule
\makecell[l]{NAE \cite{yoon2021autoencoding}\\ \\}    & \makecell[l]{Normalized autoencoder, an energy-based\\ formulation of AE.} & RE & 3.5  & $\times$ \\
\makecell[l]{GOOD \cite{bitterwolf2020certifiably}\\  \\} & \makecell[l]{Guaranteed OOD Detector, based on\\ the interval bound propagation \cite{gowal2018effectiveness}.} & MSP & 1.8& $\bigcirc$  \\
ACET \cite{Hein_2019_CVPR} & Adversarial Confidence Enhancing Training. & MSP & 1.8 & $\bigcirc$ \\
CEDA \cite{Hein_2019_CVPR} &Confidence Enhancing Data Augmentation.& MSP & 1.8 & $\bigcirc$ \\
\makecell[l]{SSD \cite{sehwag2021ssd} \\ \\} &  \makecell[l]{$k$-means clustering and mahalanobis distance \\ in the self-supervised feature space.} & Distance & 7.8 & $\times$ \\
\makecell[l]{MD \cite{lee2018simple} \\ \\ }  & \makecell[l]{Mahalanobis distance in classifier feature spaces.\\ The validation split of test OOD dataset is used \\to tune some parameters.} & Distance & 46.5 & $\bigcirc$ \\
\makecell[l]{SNGP \cite{liu2020simple} \\ \\ }  & \makecell[l]{Spectral-normalized Neural Gaussian Process\\ with bi-Lipschitz condition} &  Dempster-Shafer & 6.7 & $\times$ \\
\makecell[l]{ATOM \cite{chen2021atom}} & \makecell[l]{A classifier trained with informative outlier \\mining and adversarial training.} & Logit & 6.0  & $\bigcirc$\\
\makecell[l]{OE \cite{hendrycks2018deep}} & \makecell[l]{Outlier Exposure. A classifier trained to be \\less confident on auxiliary OOD data.} & MSP & 1.8 & $\bigcirc$ \\
\makecell[l]{ROWL \cite{sehwag2019rowl}} & \makecell[l]{A classifier trained with additional background \\classes and adversarial training.} & Logit & 5.4 & $\bigcirc$ \\
\makecell[l]{CSI \cite{tack2020csi}} & \makecell[l]{The combination of multiple OOD metrics\\ in a self-supervised feature space.} & Distance & 266.8 & $\times$ \\
\bottomrule
\end{tabular}

\end{footnotesize}
\end{center}
\vskip -10pt
\end{table*}

\section{Evaluation via Generation for OOD Detectors} \label{sec:evg}

Here, we introduce \textbf{Evaluation via Generation} for OOD Detectors (EvG), an evaluation framework for the robustness of OOD detectors under perceptually meaningful variations of OOD data. EvG consists of two modules: a variation model and an adversarial distribution. A variation model is responsible for defining the space of variation.
An adversarial distribution is a probability distribution defined on a variation model, guiding the optimization process for the worst-case failure of an OOD detector being evaluated.
After a variation model and an adversarial distribution are constructed, then MCMC sampling is performed to find an outlier sample within the variation model that is classified as in-distribution with the strongest confidence by a detector.

\subsection{Variation Models} \label{sec:variations}

A variation model is a set containing plausible variations of a test OOD dataset. 
First, we set a low-dimensional compact latent code space $\mathcal{Z}\subset \RE^{D_\bz}$.
Then, we define a variation model $\mathcal{T}_g$ using a generator $g(\bz)$, a mapping from $\mathcal{Z}$ to data space $\mathcal{X}$:
\begin{align}
    \mathcal{T}_{g} = \{\bx=g(\bz) |\; \bz \in \mathcal{Z} \}. \label{eq:variation-model}
\end{align}
Thus, $\mathcal{T}_{g}$ forms a $D_{\bz}$-dimensional manifold in the data space, reflecting the intuition that variations in perceptual stimuli may have a low-dimensional structure \cite{seung2000manifold}.
With the careful choice of $g(\bz)$, we can ensure that all elements in $\mathcal{T}_g$ are outliers and important mode of variations in a test OOD dataset is reflected.

\textbf{Instance-Conditional Variation Models.}
First, we choose $g(\bz)$ to represent identity-preserving transforms, i.e., a transformation function which does not alter the semantic content but changes the low-level characteristics of data. 
For images, the affine transform and the color jittering could be identity-preserving. Such transforms are widely used in data augmentation techniques, self-supervised learning \cite{chen2020simple}, and adversarial attack \cite{chen2020explore,yang2021random}, but new in OOD detection. To transform, $g$ needs a base data as an input.
\begin{align}
    \bx = g(\bz; \bx_{out}),\quad \bx_{out} \in \mathcal{D}_{out}.
\end{align}
Then, we obtain an \textbf{instance-conditional variation model} $\mathcal{T}_{g}(\bx_{out})$ from Eq.~(\ref{eq:variation-model}).
In our experiments with image data, we define \textbf{Affine} variation model ($D_{\bz}=5$), which has rotation, translation (x, y), scaling, and shear (x, y) operations, and \textbf{Color} variation model ($D_{\bz}=4$), which can adjust brightness, contrast, saturation and hue of an image. Both are implemented by TorchVision.

\textbf{Unconditional Variation Models.}
A novel outlier example can be synthesized by re-combining the existing features in the test OOD dataset.
Synthesizing a new example with a high fidelity is generally a difficult task, but recent developments in generative modeling provide strong tools, such as StyleGAN2 \cite{Karras_2020_CVPR}.
We use the generator of StyleGAN2 as $g(\bz)$.
This results in an unconditional variation model $\mathcal{T}_{g}$.
In our experiments, we obtain $g(\bz)$ by training StyleGAN2 on the test OOD set $\mathcal{D}_{out}$ and refer the resulting variation model as \textbf{GAN} variation model. In this case, the domain of $\bz$ is the surface of a unit sphere in 64-dimensional space.

\begin{figure*}
    \centering
    \includegraphics[width=0.45\textwidth]{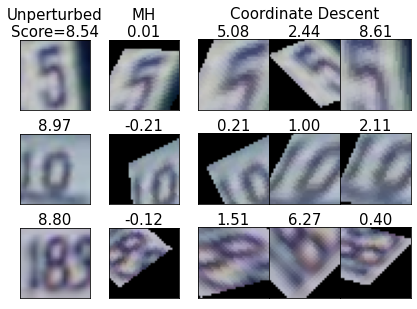}
    \includegraphics[width=0.45\textwidth]{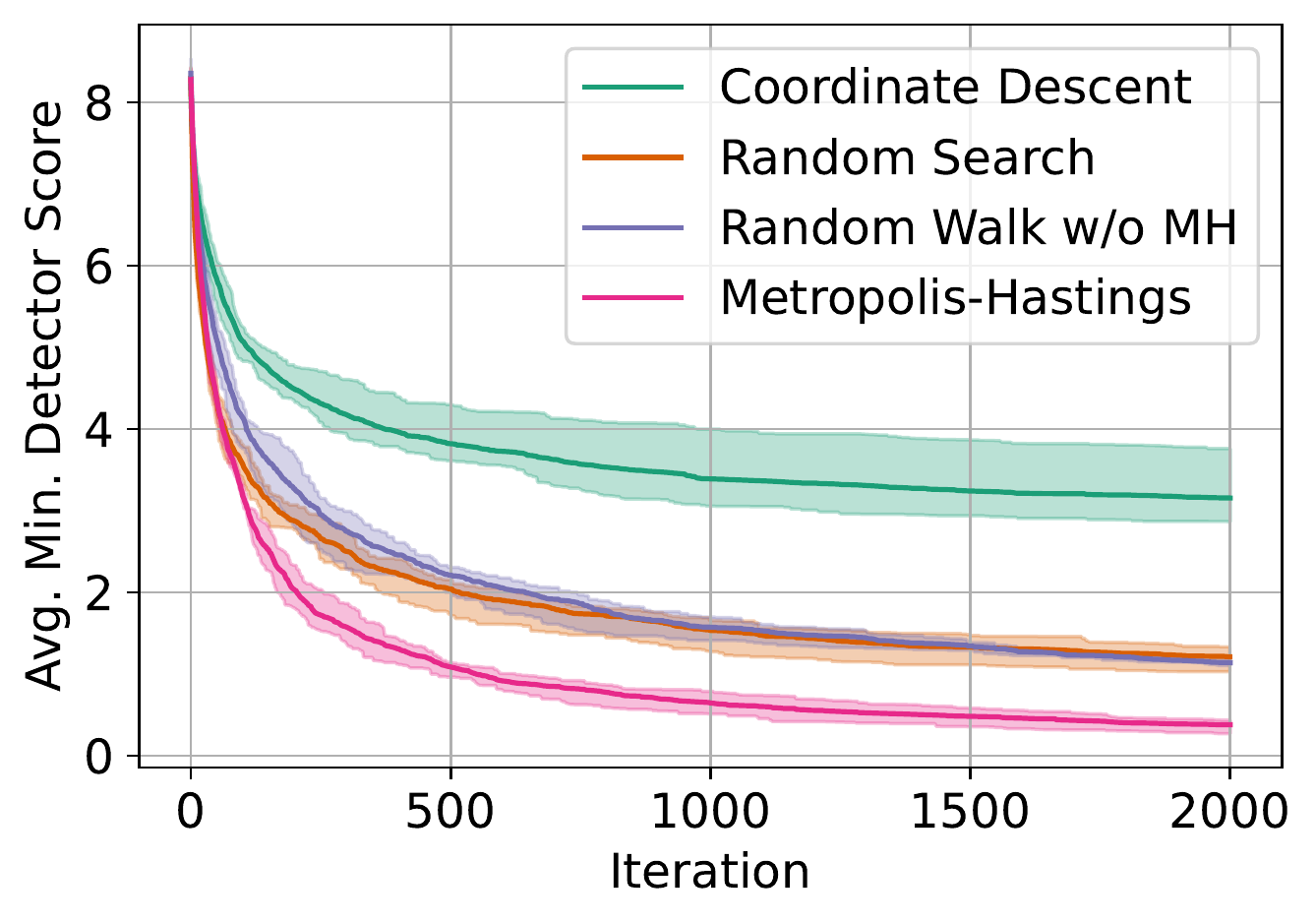}
    \caption{
    An OOD detector $f$ (CIFAR10 images as inliers) is investigated under the affine transform variation model with the SVHN images.
    (Left) Our sampling-based method, MH, finds one global solution unlike the coordinate descent method.
    (Right) Average of the minimum detector scores over 64 different variation models, and the shaded interval that represents the max and min values. 
    }
    \label{fig:local-optima}
\vskip -5pt
\end{figure*}

\subsection{Adversarial Distributions} \label{sec:advdist}

With a variation model $\mathcal{T}$, our goal is to find the worst-case example via solving the following optimization problem with respect to the latent code $\bz$:
\begin{align}
    \bz^* = \argmin _{\bz \in \mathcal{Z}} f(g(\bz)) ; \quad \bx^* = g(\bz). \label{eq:advsearch}
\end{align}
where $\bx^*$ is an outlier that the detector considers being the most inlier in $\mathcal{T}$. 
Instead of directly solving Eq.~(\ref{eq:advsearch}), we define a probability distribution that can guide our optimization similarly to simulated annealing \cite{pincus1970letter, kirkpatrick1983optimization}.

An \textbf{adversarial distribution} against an OOD detector $f$ on a variation model $\mathcal{T}$ is defined as follows:
\begin{align}
    p_f(\bz) = \frac{1}{Z}\exp(- f(g(\bz))/T), \quad Z=\int_{\mathcal{Z}} \exp(-f(g(\bz))/T) d\bz  <\infty \label{eq:advdist}
\end{align}
where $T>0$ is the temperature.
The form of $p_f(\bz)$ in Eq.~(\ref{eq:advdist}) can be seen as an instance of a Gibbs distribution or an energy-based model. Here, the detector score $f(\bx)$ serves as an energy. 
Note that when $\mathcal{T}_g$ is instance-conditional, then there is an adversarial distribution for each test OOD datum. 

An adversarial distribution $p_f(\bx)$ has two important properties by construction.
First, by construction, an adversarial distribution assigns a high probability density on samples likely to be classified as in-distribution by the detector.
Second, for the limit $T\to0$, sampling from an adversarial distribution becomes equivalent to the adversarial search, as the probability mass is concentrated on $\bx^*$ in Eq.~(\ref{eq:advsearch}).
Hence, sampling from an adversarial distribution can be viewed as a relaxation of the original optimization problem Eq.~(\ref{eq:advsearch}), and $T$ governs the degree of relaxation.

\subsection{MCMC for Optimization}
We solve the original optimization problem in Eq.~(\ref{eq:advsearch}) by sampling from an adversarial distribution. 
Markov Chain Monte Carlo (MCMC) is employed to draw samples from an adversarial distribution. 
Due to the ergodicity of a Markov chain, the chain is guaranteed to visit every minima of the detector score $f(\bx)$ given a sufficient number of steps.
By choosing the sample with the lowest detector score from the whole trajectory, we can find the global minima of $f(\bx)$.
In this way, we are more likely to find the global minima compared to local optimization methods since the chain in MCMC explores the entire search space.
While any MCMC algorithm can be applicable, we employ Metropolis-Hastings algorithm (MH, \cite{metropolis1953equation}), because MH algorithm is simple, robust, easy to implement, and has a small number of hyperparameters.

In EvG, an adversarial distribution $p_f(\bz)$ and a generator $g(\bz)$ in fact form a generative model over $\mathcal{T}_g$.
To draw sample from EvG's generative model we perform MCMC, which will result in the optimization of Eq.~(\ref{eq:advsearch}) in the limit.

Figure \ref{fig:local-optima} shows that sampling approach is indeed effective in optimization, whereas a local optimization algorithm, the coordinate descent in this case,  would stuck in local optima.
This is consistently observed throughout many detector models, which seems because a variation model $\mathcal{T}$ contains perceptually diverse samples, and a detector function $f$ is often very non-convex and non-linear.

\begin{table*}
  \caption{Evaluation of OOD detectors using the adversarial distributions. Clean indicates the test split of a test OOD dataset. AUC scores are evaluated using 10,000 inliers and 5,000 outliers. MinRank is computed from a run which consists of 1,000 MCMC chains. The standard error is computed from five independent runs. The boldface are the largest numbers and the underlined are the smallest numbers among strong detectors.}
  \setlength\tabcolsep{1pt} 
  \label{tb:perf}
  \begin{center}
  \begin{footnotesize}
  \begin{tabular}{lccccc|ccccc|cc|cc}
    \toprule
    Metric & \multicolumn{10}{c|}{AUC} & \multicolumn{4}{c}{MinRank} \\
    \midrule
    OOD & \multicolumn{5}{c}{SVHN}  & \multicolumn{5}{c|}{CelebA} & \multicolumn{2}{c}{SVHN} & \multicolumn{2}{c}{CelebA}  \\
    \midrule
    Variation & Clean & $l_\infty$ & Affine & Color & GAN  & Clean & $l_\infty$ & Affine & Color & GAN & Clean & GAN & Clean & GAN \\
    \midrule
    \multicolumn{3}{l}{\textbf{Weak Detectors}} \\
Glow & .069 & .071 & .002 & .000 & .000 & .542 & .547 & .008 & .001 & .036 & 8 & 1 $\pm$ 0 & 177 & 24 $\pm$ 1\\
PixelCNN++ & .076 & .078 & .015 & .000 & .000 & .639 & .644 & .085 & .003 & .053 & 0 & 0 $\pm$ 0 & 391 & 28 $\pm$ 4\\
AE & .080 & .054 & .032 & .000 & .000 & .533 & .401 & .159 & .001 & .012 & 0 & 0 $\pm$ 0 & 18 & 1 $\pm$ 0\\
    \midrule   \multicolumn{3}{l}{\textbf{Strong Detectors}} \\ 
NAE & \underline{.935} & .900 & .755 & .704 & \underline{.005} & .874 & .803 & .627 & .343 & .098 & \underline{11} & \underline{0} $\pm$ 0 & 616 & 39 $\pm$ 4\\
GOOD & .943 & .714 & \underline{.565} & .664 & .515 & .939 & .852 & .580 & .664 & .570 & 2141 & 1609 $\pm$ 85 & 3558 & 1720 $\pm$ 161\\
ACET & .966 & .905 & .753 & .868 & .425 & .986 & .952 & .593 & .904 & .631 & 3414 & 1304 $\pm$ 74 & 4845 & 2917 $\pm$ 158\\
CEDA & .987 & .690 & .636 & .912 & .435 & .981 & .878 & .343 & .908 & .636 & 3348 & 0 $\pm$ 0 & 3200 & \underline{0} $\pm$ 0\\
SSD & .989 & .199 & .631 & .936 & .479 & \underline{.780} & .231 & .564 & .674 & .288 & 4133 & 691 $\pm$ 54 & 2335 & 1211 $\pm$ 56\\
MD & .993 & .581 & .747 & \underline{.484} & .169 & .796 & .035 & .394 & \underline{.064} & \underline{.046} & 69 & 2 $\pm$ 0 & \underline{5} & 1 $\pm$ 0\\
SNGP & .994 & \underline{.096} & .761 & .892 & .488 & .882 & \underline{.007} & \underline{.336} & .613 & .145 & 2427 & 0 $\pm$ 0 & 8 & 13 $\pm$ 2\\
ATOM & .996 & \textbf{.992} & .908 & .974 & .616 & \textbf{.998} & \textbf{.992} & .867 & \textbf{.976} & \textbf{.771} & 6203 & 1522 $\pm$ 98 & \textbf{7517} & 1240 $\pm$ 298\\
OE & .997 & .566 & .751 & .953 & .620 & .992 & .593 & .678 & .916 & .736 & 5636 & 1774 $\pm$ 102 & 5217 & 1947 $\pm$ 161\\
ROWL & .997 & .875 & .675 & .969 & .659 & .991 & .809 & .374 & .933 & .559 & 5213 & 260 $\pm$ 179 & 2454 & 644 $\pm$ 100\\
CSI & \textbf{.998} & .630 & \textbf{.940} & \textbf{.992} & \textbf{.893} & .890 & .463 & \textbf{.892} & .855 & .484 & \textbf{8208} & \textbf{6866} $\pm$ 112 & 3727 & \textbf{3225} $\pm$ 53\\

    \bottomrule
  \end{tabular}
  \end{footnotesize}
  \end{center}
  \vskip -15pt
\end{table*}

\section{Experiments}  \label{sec:experiments}

In our experiments, we aim to evaluate the robustness of the state-of-the-art OOD detectors under variation models.
Using Affine, Color, and StyleGAN2 variation models, adversarial distributions are constructed for 13 OOD detectors, described in Table \ref{tb:ood-detectors}. 
We first show that adversarial distributions can find the known weaknesses of baseline detectors.
Then, we apply our methods on the state-of-the-art OOD detectors which show near-perfect performance on a popular benchmark.
Finally, we reveal some interesting patterns in the failure modes of the detectors.
Table \ref{tb:perf} summarizes our experimental results.
The extended experimental results can be found in Appendix.

\subsection{Experimental Settings}
\label{ssec:exprerimental settings}

\textbf{Datasets.} 
We use CIFAR-10 as in-distribution data $\mathcal{D}_{in}$. SVHN and CelebA are used as test OOD datasets $\mathcal{D}_{out}$.
All datasets are splitted into training, validation, and testing sets.
All data are 32$\times$32 RGB images. Details on datasets can be found in Appendix.

\begin{wrapfigure}{r}{0.43\textwidth}
  \vskip -10pt
  \centering
  \includegraphics[width=0.4\textwidth]{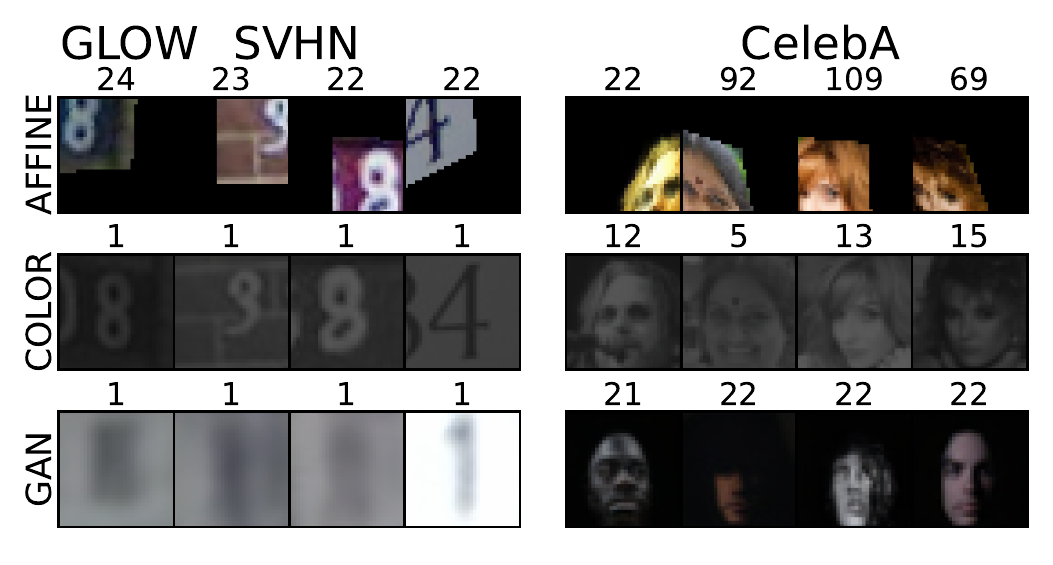}
  \includegraphics[width=0.4\textwidth]{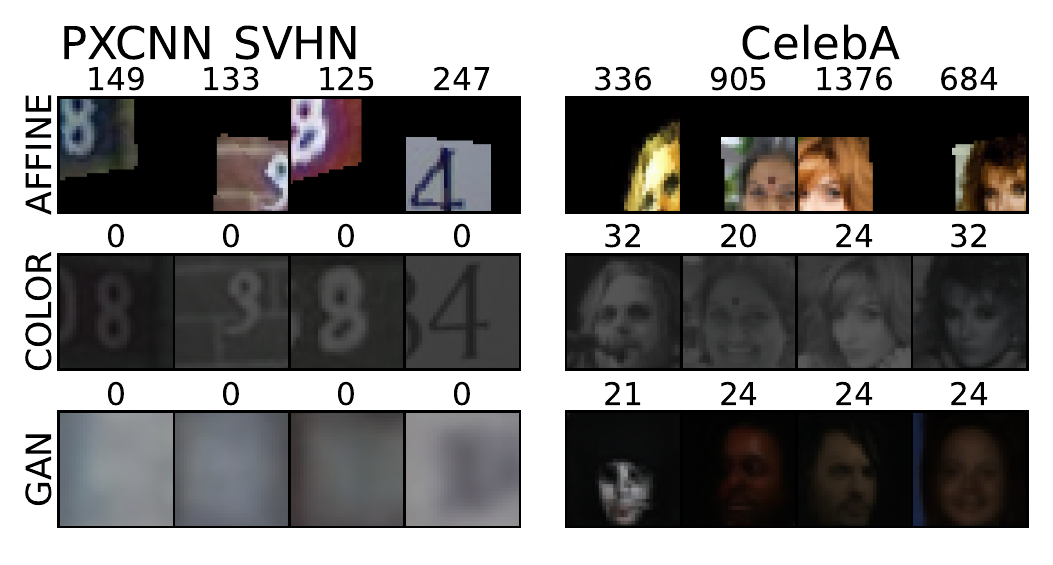}
  \includegraphics[width=0.4\textwidth]{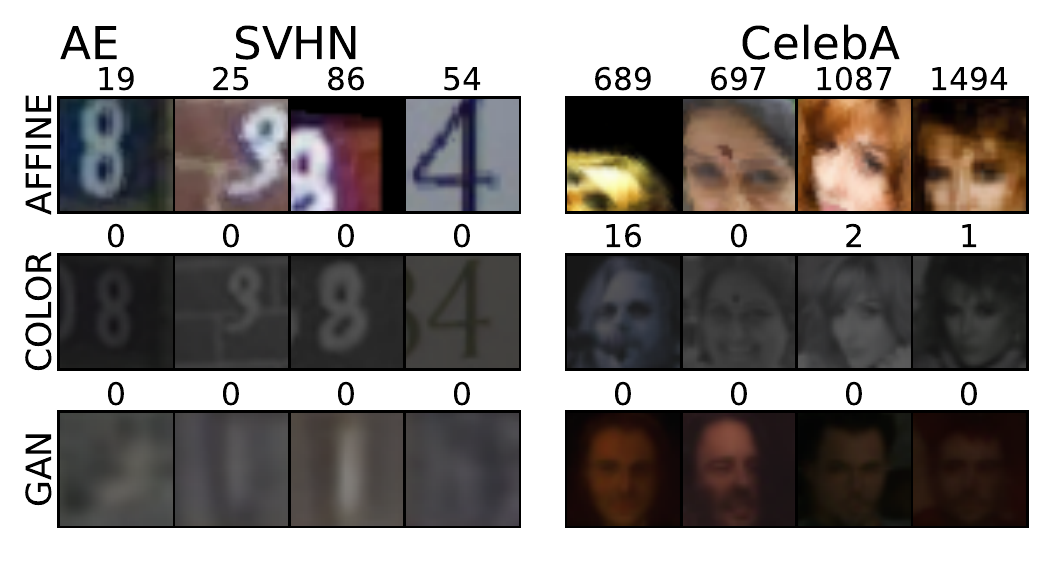}
  \vskip -10pt
  \caption{
  The worst-case samples for the weak detectors. The numbers are the detector score rank.   } \label{fig:weak} 
  \vskip -10pt
\end{wrapfigure}

CIFAR-10 (in) vs SVHN (out) dataset pairs become popular after the reports showing that OOD detectors built upon deep generative models, such as PixelCNN++ or Glow, dramatically fail to detect SVHN as outliers \cite{hendrycks2018deep,nalisnick2018do}, producing AUC scores close to 0.
After this observation, multiple OOD detectors achieving AUC scores higher than 0.9 or even higher than 0.99 on the benchmark have been proposed (see Table \ref{tb:perf}).
Given their near-perfect detection score, we question whether the detectors are indeed robust OOD detectors under variabilities likely to occur for image data.

\textbf{OOD Detectors.} 
We implement 13 previously proposed outlier detectors, listed in Table \ref{tb:ood-detectors}.
The detectors are grouped into two, the weak and the strong, based on their performance on CIFAR-10 vs. SVHN.
The \emph{weak} detectors are outlier detectors show AUC score close to zero on CIFAR-10 vs. SVHN.
\emph{Strong} detectors are selected from three criteria: 1) AUC score on CIFAR-10 vs SVHN is higher than 0.9. 2) The code is publicly available and written in PyTorch. The language constraint is introduced for the unified experiment. 3) The performance claimed in the original work should be reproducible.

Extended experimental results with more OOD detectors can be found in Appendix.
OOD detectors we considered but not included in the experiment of this paper are also discussed in Appendix. 

\textbf{Evaluation Metrics.}
For each detector, we report AUC score of classifying the set of outliers and the test split of $\mathcal{D}_{in}$.
AUC score close to 1.0 indicates a successful OOD detection.
Additionally, for unconditional adversarial distributions, we also compute \textbf{MinRank}, the rank of the smallest detector score achieved. The rank of the score is computed with respect to the scores of the test split of $\mathcal{D}_{in}$, which contains 10,000 samples. An ideal OOD detector should score MinRank of 10,000, since a rank starts from zero.

\subsection{Implementation of EvG}
\label{ssec:imp.advdist}

\textbf{Variation Models.}
We use Affine ($D_\bz=5$), Color ($D_\bz=4$), and GAN ($D_\bz=64$) variation models. The first two are instance-conditional, and GAN is an unconditional variation model.
Variation models are constructed for each test OOD dataset, SVHN and CelebA
For StyleGAN2, we use an implementation provided by PyTorch-StudioGAN repository\footnote{https://github.com/POSTECH-CVLab/PyTorch-StudioGAN} \cite{kang2020ContraGAN}.

\textbf{Adversarial Distributions.} 
To deal with the scale difference in detector scores $f(\bx)$, the scores are standardized to have the zero mean and the variance of one, when evaluated on the validation split of in-distribution data.
For all experiments, the temperature is set $T=1$.

MH algorithm with Gaussian proposal distribution is used. The standard deviation of the proposal distribution is fixed to $0.1$. 
A proposal is randomly accepted based on Metropolis' criterion. We run 5,000 independent Markov chains with each chain runs for 2,000 steps. 
For instance-conditional adversarial distributions, we use the first 5,000 examples of a test OOD dataset.
Among the visited states in a trajectory, the sample with the smallest $f(\bx)$ is selected.
More implementation details can be found in Appendix.

\textbf{$l_\infty$ Attack.}
For comparison, we also implement a black-box $l_\infty$ attack based on the zeroth-order random coordinate descent with momentum \cite{chen2017zoo} and run for each OOD detectors for 20,000 steps.
Following \cite{bitterwolf2020certifiably}, we perform the attack on a ball of radius 0.01.
The momentum coefficient was 0.999, and the step size starts from 0.1 and is halved every 2,000 steps.
Note that our black-box $l_\infty$ should be weaker than their white-box version used in \cite{bitterwolf2020certifiably}.

\begin{figure*}
    \centering
    \includegraphics[width=0.9\textwidth]{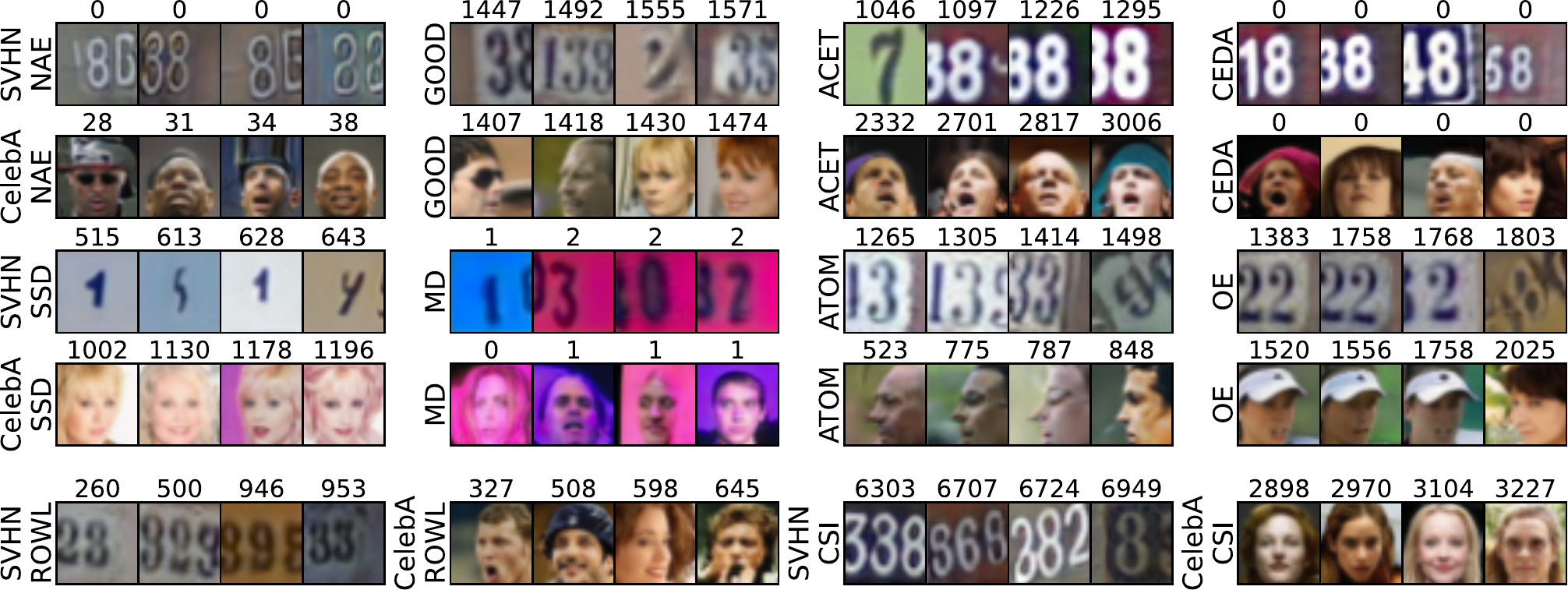}
    \caption{The worst-case samples from unconditional adversarial distributions against 10 strong detectors under GAN variation model on SVHN and CelebA. The numbers indicate the detector score rank in-distribution test set.}
    \label{fig:stylegan2}
    \vskip -10pt
\end{figure*}

\subsection{Adversarial Distributions Against Detectors with Known Weaknesses} \label{ssec:weak detector}

We confirm the effectiveness of adversarial distributions by applying them on the weak detectors, Glow, PXCNN, and AE, where the weaknesses have already been analyzed.
It is known that Glow, PXCNN, and AE erroneously classify low-complexity images, such as monotone images or highly blurry images, as in-distribution \cite{Serra2020Input,yoon2021autoencoding}.
The known failure mode can be clearly observed from Figure \ref{fig:weak}.
The worst-case samples from adversarial distributions are consistent with the previously known tendency. The samples are very blurry and a large portion of pixels are monotone, which are misclassified as in-distribution by the weak detectors with high confidence.

\subsection{Evaluating State-of-the-Art Detectors} \label{ssec:strong detector}

Then, we evaluate the robustness of 10 strong detectors using the proposed variation models and adversarial distributions.
Table \ref{tb:perf} provides the quantitative evaluation results. Figure \ref{fig:stylegan2} show examples from each adversarial distributions.
As a whole, it is difficult to state a winner, as no detector is perfectly robust. All the tested detectors demonstrate notable drops in detection performance under more than one types evaluation set-ups.
Additionally, our method reveals interesting, previously unexplored blind spots of OOD detectors.
Due to the space constraint, more results can be found in Appendix.

\textbf{Effectiveness of Low-Dimensional Variation.} 
Considering their low dimensionality, Affine and Color variation models are surprisingly effective at finding failure modes for some models. 
GOOD, ACET, and CEDA, the detectors trained to be robust against $l_\infty$ threat model, show significant degree of vulnerability under Affine.
This indicates that optimizing for the robustness against one threat model does not necessarily improve the robustness against other threat models.
Meanwhile, CSI show strong robustness against Affine and Color, probably because transformations similar to Affine and Color are used during its training.

\textbf{Characteristic Failure Modes.} From Figure \ref{fig:stylegan2}, there are several noticeable patterns in the failure modes of the detectors.
First, some OOD detectors exhibit the \textbf{color bias}, classifying an image of a certain color as in-distribution. For example, MD shows a strong bias by confidently classifying images with pink or cyan color as in-distribution, recording the lowest AUC under Color on both datasets.

In Figure \ref{fig:stylegan2}, each OOD detector has different blind spots, showing distinct failure examples. ACET tends to classify generated CelebA images with a certain facial expression as inliers.
SSD consistently misclasifies photos of a white blond female.

\begin{wrapfigure}{r}{0.4\textwidth}
  \vskip -0.5in
  \centering
  \includegraphics[width=0.38\textwidth]{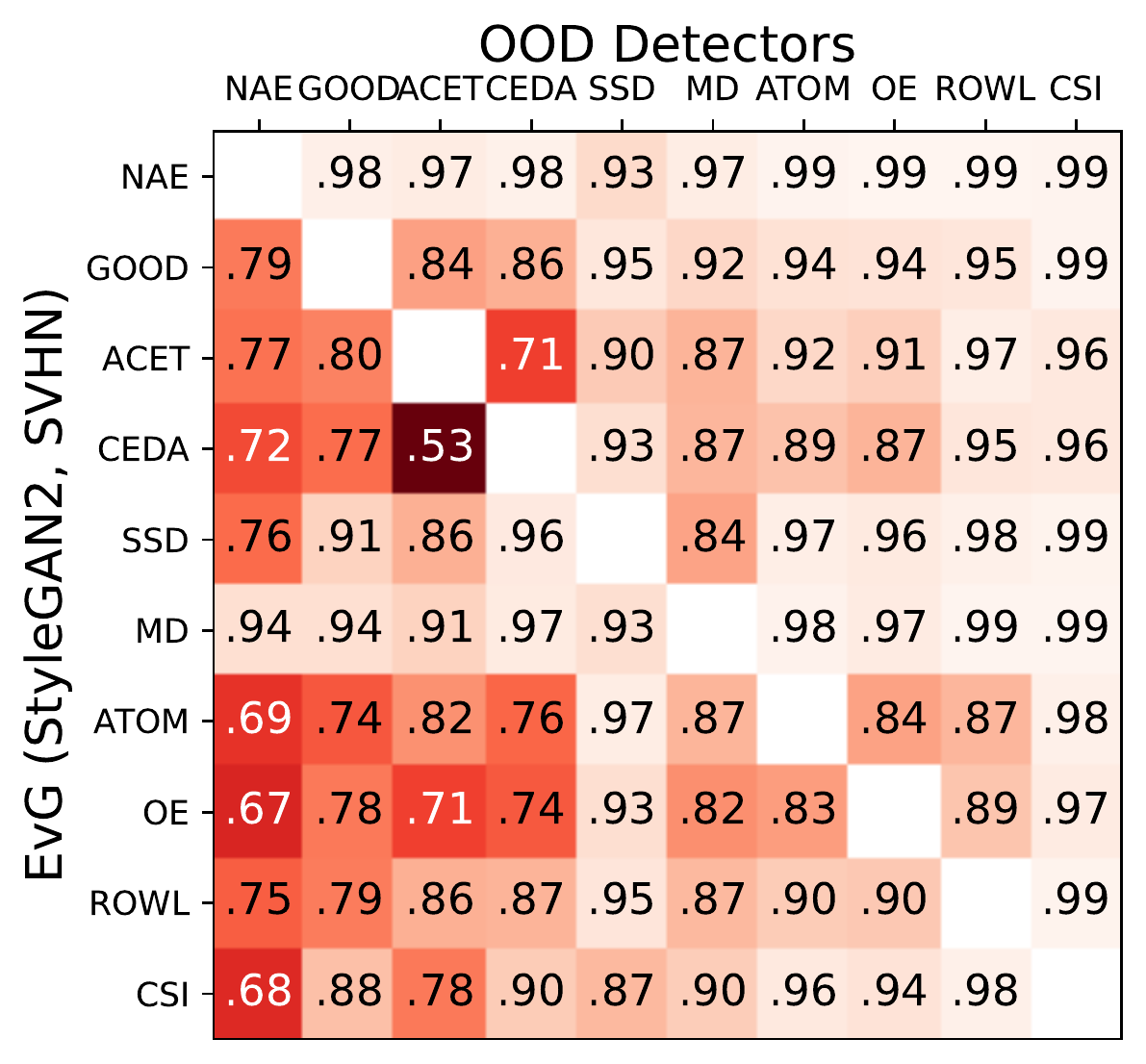}
  \includegraphics[width=0.38\textwidth]{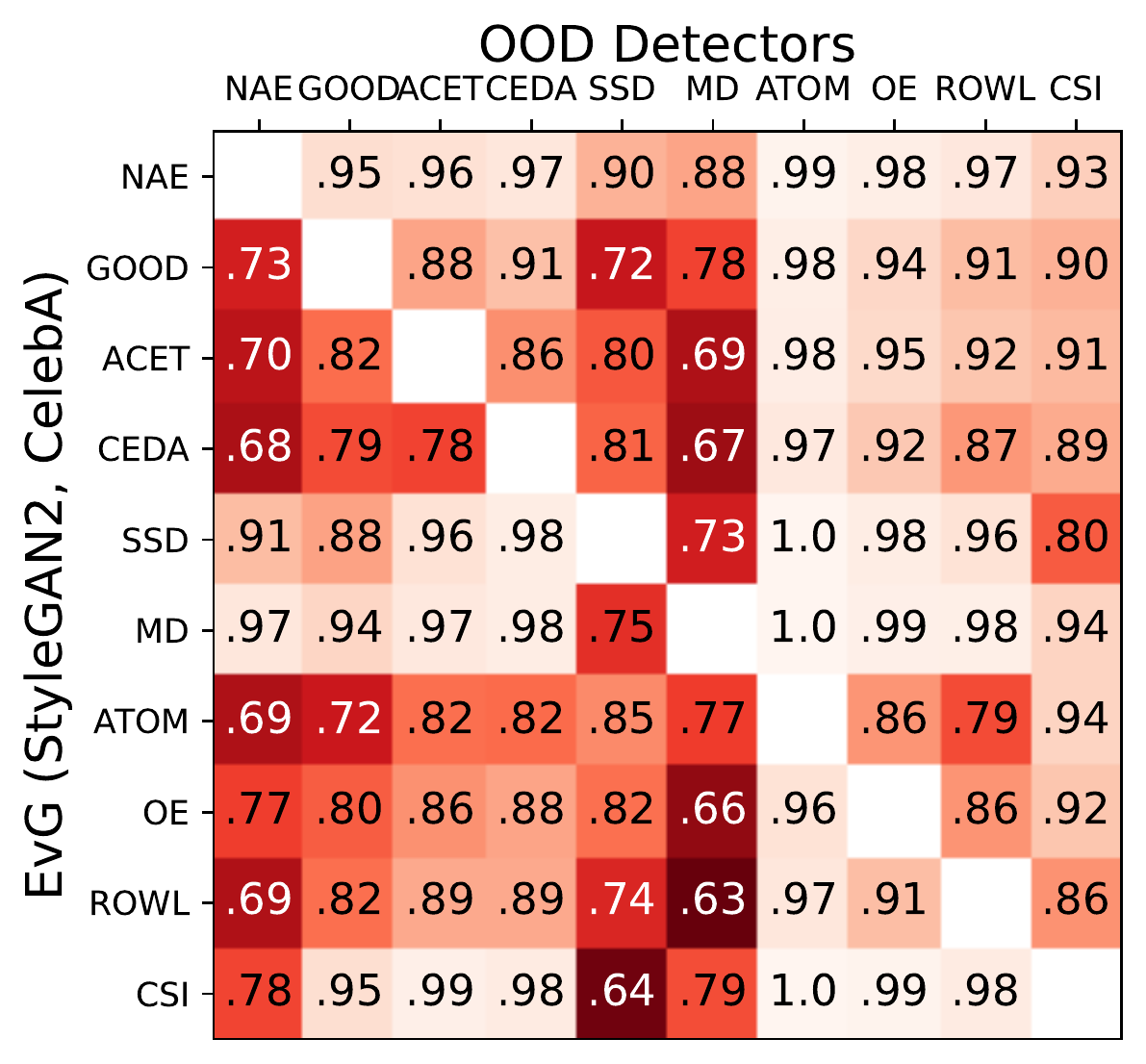}
  \vskip -10pt
  \caption{ AUC scores evaluated against other detectors' adversarial distribution samples. (up) SVHN (down) CelebA } \label{fig:pairwise} 
  \vskip -0.6in
\end{wrapfigure}

\textbf{Transferability of Failure Modes.}
We question whether samples from an adversarial distribution is transferable, i.e., given two detectors $f_1(\bx)$ and $f_2(\bx)$, are the worst-case outliers from $p_{f_1}(\bx)$ able to deceive another detector $f_2(\bx)$?
We collect samples from $p_{f_1}(\bx)$ then check if $f_2(\bx)$ can detect them as outliers. The pairwise results are displayed in Figure \ref{fig:pairwise}.

First, some of the failure modes are shared. Worst-case outliers from other detectors are able to fool NAE and MD, indicating that the outliers are also one of the failure modes of them.
Second, the failure mode of one detector is more readily transferable to another detector with the similar inductive bias. For example, GOOD, ACET, CEDA, and OE share the same structure as a classifier. Both SSD and CSI utilize the self-supervised learning. 
Third, the failure modes are not very transferable between strong detectors built upon different inductive biases, such as OE and CSI. They are successful in detecting each other's worst-case samples.

\section{Discussion}
\label{sec:discussion}

\textbf{Extensions.}
As a framework, the proposed method has a significant degree of flexibility in designing a variation model $\mathcal{T}_{g}$.
Therefore, with an appropriate choice of $g(\bz)$, our method is certainly extensible to other types of data, such as audio anomaly detection \cite{koizumi2019toyadmos} or robot collision detection \cite{park2021collision}.
Our method will be particularly helpful in a domain where additional collection of test OOD data is infeasible or costly.
Also, our method will be directly benefited from the rapid progress in generative modelling. The fidelity and coverage of our protocol would increase as generative models become more capable of generating diverse and realistic data.

\textbf{Future Work.}
To facilitate the development of OOD detection methods, we intend to publish EvG as an online software suite for evaluating the robustness of OOD detectors.  
We also plan to publish an online leaderboard displaying the results from an adversarial distribution analysis, and to update the leaderboard with the latest OOD detection methods.

\section{Conclusion}
In this paper, we have addressed the limitations of the current robustness evaluation protocol and proposed a novel framework, Evaluation via Generation, which utilize a generative approach to investigate the failure modes of OOD detectors beyond noise-like perturbations or preset corruptions.
The proposed framework posing new challenges in OOD detection by discovering unexplored weaknesses in the existing OOD detectors, 
We expect our framework to stimulate the advance of trustworthy machine learning, where a model needs to be evaluated beyond the fixed test dataset.

\clearpage
\small
\bibliographystyle{unsrt}
\bibliography{ref}


\clearpage

\appendix

\section*{Appendix}

Appendix is organized as follows:
\begin{itemize}[leftmargin=1em,topsep=0pt,noitemsep]
\item Section A: Extended Experimental Results
\item Section B: Additional Discussion
\item Section C: Datasets
\item Section D: Implementation of EvG
\item Section E: Implementation of OOD Detectors
\end{itemize}

\section{Extended Experimental Results}

\begin{figure}[h]
    \centering
    \includegraphics[width=0.8\textwidth]{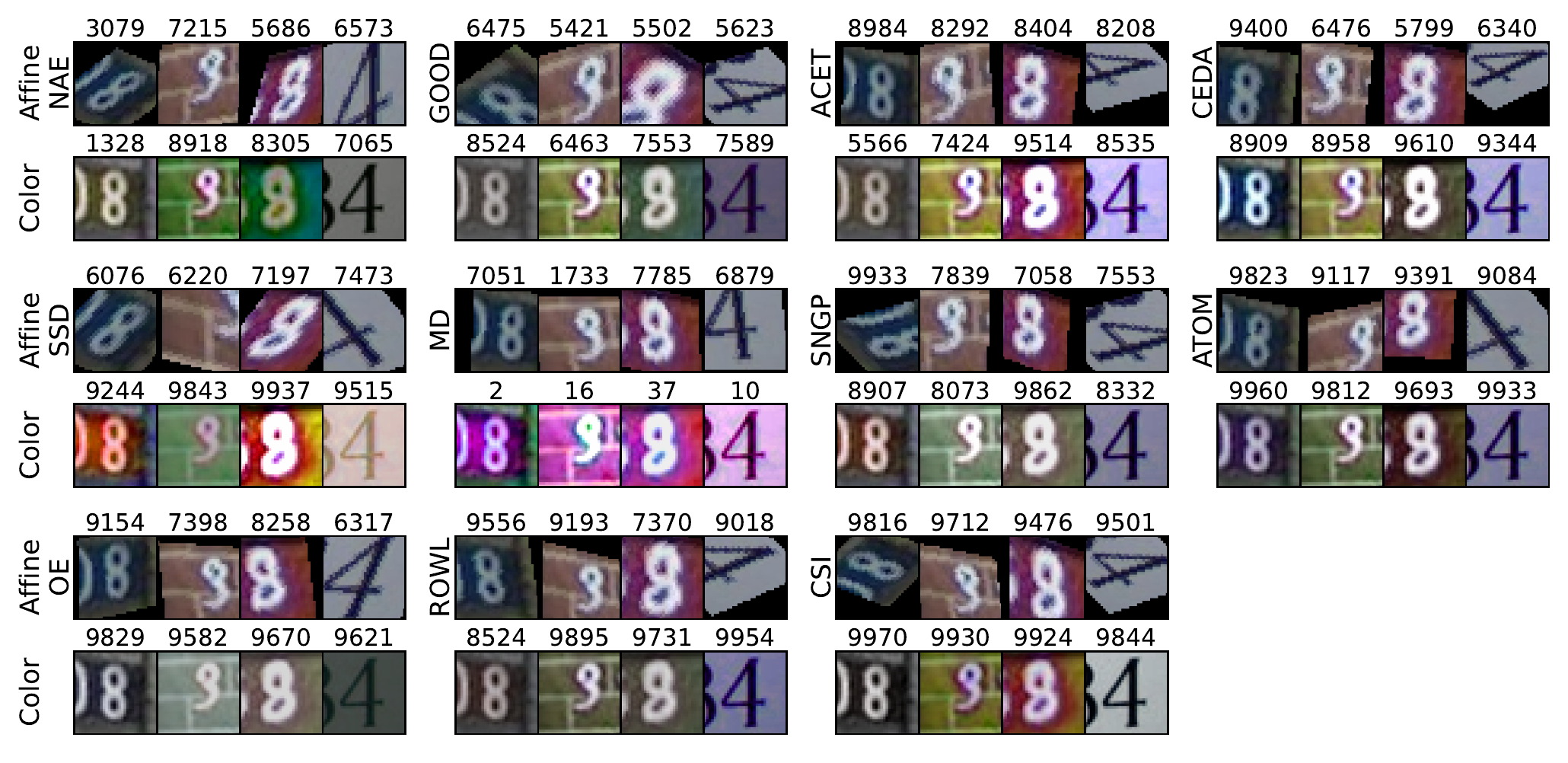}
    \caption{Examples from instance-conditional adversarial distributions against strong detectors under Affine and Color variation models on SVHN. The numbers indicate the detector score rank in-distribution test set. Note that these are not the worst-case samples but chosen to compare the results from different detectors.}
    \label{fig:svhn_strong_affine_color}
    \vskip -10pt
\end{figure}

\begin{figure}[h]
    \centering
    \includegraphics[width=0.8\textwidth]{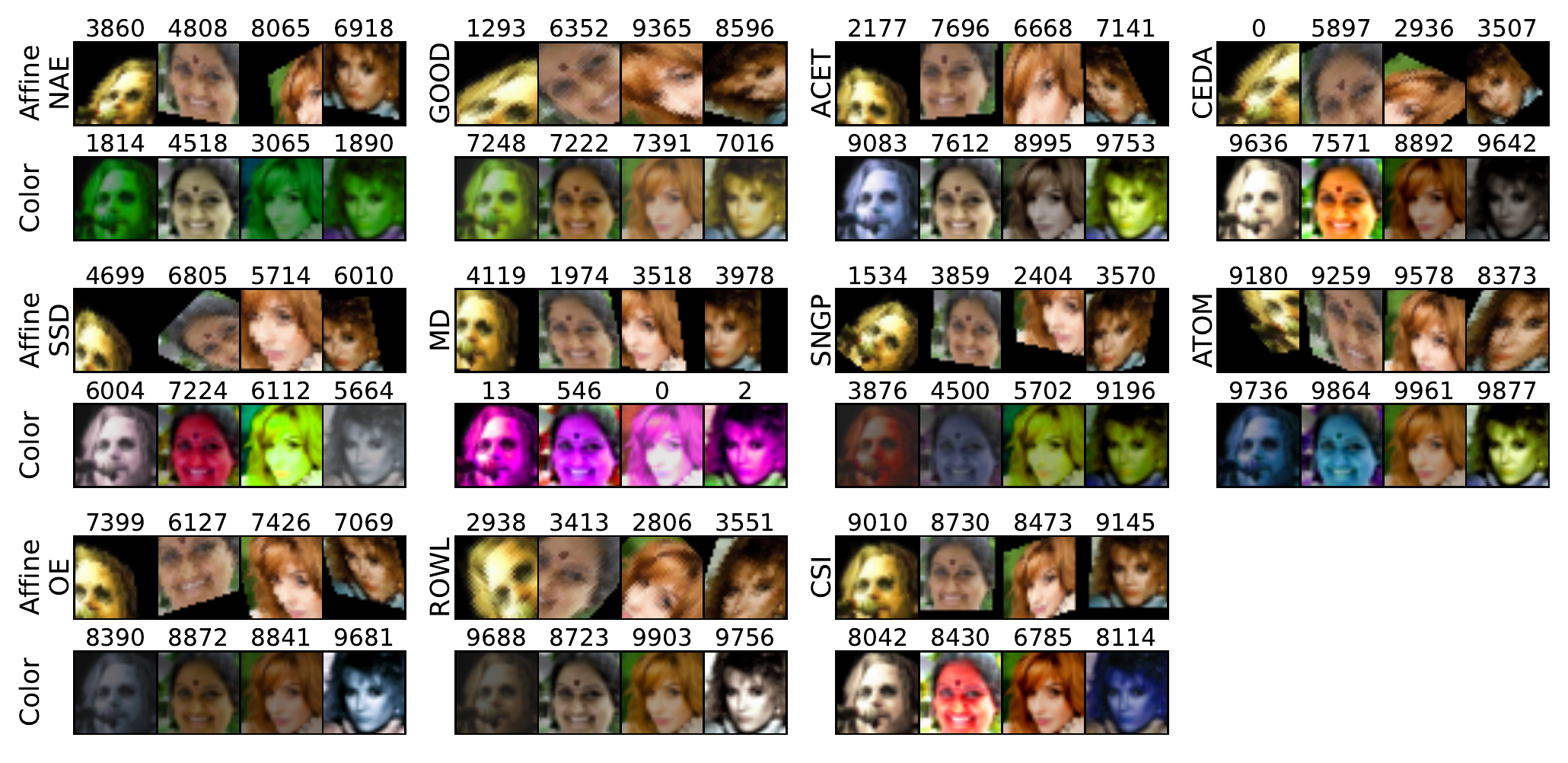}
    \caption{Examples from instance-conditional adversarial distributions against strong detectors under Affine and Color variation models on CelebA. The numbers indicate the detector score rank in-distribution test set. Note that these are not the worst-case samples but chosen to compare the results from different detectors.}
    \label{fig:celeba_strong_affine_color}
    \vskip -10pt
\end{figure}

Figure \ref{fig:svhn_strong_affine_color} and Figure \ref{fig:celeba_strong_affine_color} show the samples from instance-conditional adversarial distributions, confirming that OOD detectors can be easily fooled by simple affine and color transforms. 

\clearpage
\section{Additional Discussion} \label{appendix:discussion}

\subsection{Ethics Statement}
Our main ethical concern is that a subset of OOD detectors used in our experiment, OE, CEDA, ACET and GOOD, are trained using 80 Million Tiny Images dataset \citep{torralba200880}, which is retracted by authors over ethical concerns.
While we were aware of the issue of the dataset, the use of models trained on the dataset was inevitable because of the reproducibility.
We have tried to train OOD detectors using alternative datasets but failed to reach the performance of detectors originally trained on 80 Million Tiny Images.
To minimize the effect of the retracted dataset, the dataset was never used directly. We only used the publicly available model checkpoints, and did not download or access to a copy of dataset.

\subsection{Limitations}
Our methods still rely on a specific choice of test OOD datasets which need human design.
However, this is a limitation that is common for any existing evaluation protocols for OOD detection.
It is a fundamental challenge in OOD detection that the set of all possible outliers is so vast. Even though an OOD detector turns out to be perfect under EvG, we can not guarantee that the tested model is indeed an ideal OOD detector that can detect all possible outliers.

\subsection{Definition of Out-of-Distribution (OOD)}
Besides the definition of OOD-ness we introduced in Section 2, there is another popular definition of OOD-ness based on a density sub-level set~\cite{steinwart2005classification}.
For an $\eta>0$, $\{\bx|\; p_{in}(\bx)\leq \eta \}$ is defined as the set of all possible outliers $\mathcal{S}_{out}$. 
Setting $\eta=0$, the density sub-level set based definition coincides with the support-based definition. 
While a density sub-level set can be arbitrarily distorted under an invertible coordinate transform \citep{lan2020perfect}, the support-based definition provides an invariant characterization of OOD-ness.

\subsection{Room for Improvement in OOD Detection}

Table \ref{tb:perf} reveals that there is a room for improvement in OOD detection, even besides the samples found by EvG. First, the performance on SVHN does not transfer to CelebA. CSI, SNGP, NAE, MD, and SSD show unsatisfactory performance on CelebA. This is why using multiple OOD dataset in evaluation is important.
Second, MinRank shows that the worst-case performance of OOD detectors are not very satisfactory. Having Clean MinRank less than 5,000 means that if we set a decision threshold so that no outlier is present, we need to misclassify more than half of inliers as OOD.

\subsection{GAN Generation Failures}

We found that GAN may fail to generate a realistic image and produce a noise-like image instead. This failure does not happen when GAN is run to generate samples freely in a conventional way as a typical GAN. This phenomenon indicates that there are points in GAN's latent space that does not map to realistic images. We could not find a measure to suppress this generation failure. However, even though the generated noise are not realistic, we can tell they were not in-distribution, so we decide to keep them in the experiments.

\subsection{What Makes a Detector Robust?}

From Table \ref{tb:perf}, we can now assess the effectiveness of different ideas in robust OOD detection. First, adversarial training is effective in defending $l_\infty$ attack. In other words, an algorithm is highly unlikely to be robust against $l_\infty$ attack when it is not trained in an adversarial way. GOOD, ACET, CEDA, ROWL, and ATOM are all trained adversarially and show better $l_\infty$ robustness than the others. 
Second, many OOD detectors rely on auxiliary dataset. Given that no auxiliary data is used, the performance of CSI is impressive.

\section{Datasets} \label{appendix:datasets}

In our experiments, we use CIFAR-10, SVHN, and CelebA to demonstrate the proposed methods and to evaluate OOD detection algorithms, and their details are described in Table \ref{tb:dataset}.
In all cases, we use the official test split as our test datasets.
For CIFAR-10 and SVHN, the official train split is randomly splitted into train (90\%) and valid split (10\%).
For CelebA, we use the official train-valid split as it is given. Each CelebA image is center-cropped into a 140$\times$140 image and scaled to 32$\times$32 using the bilinear interpolation.

\begin{table}[h]
    \centering
    \caption{Statistics for datasets.}
    \vspace{3pt}
    \label{tb:dataset}
    \begin{tabular}{ccccc}
         Dataset & License & Train & Valid & Test \\ 
         \midrule
         CIFAR-10 \citep{krizhevsky2009learning}& Not described & 45,000 &5,000 & 10,000 \\
         SVHN \citep{netzer2011reading}& For non-commercial use only  & 65,930 & 7,327 & 26,032 \\
         CelebA \citep{liu2015faceattributes}& For non-commercial research purposes only & 162,770 & 19,867 & 19,962 \\ 
    \end{tabular}
\end{table}

No augmentation is applied to test data. However, random horizontal flip with probability 50\% and uniform dequantization \citep{theis2015note} are applied to training images in the training of the binary classifiers and the autoencoders in the adversarial search and adversarial distributions.

\section{Implementation of EvG} \label{sec:app-impl}

\subsection{Metropolis-Hasting Algorithm} \label{ssec:mh}

To generate samples from an adversarial distribution, we apply Metropolis-Hastings algorithm \citep{metropolis1953equation} in $\mathcal{Z}$.
An initial state $\bz_0$ is drawn from an uniform distribution.
Given a state $\bz_t$ at time $t$, a candidate for the next state $\bar{\bz}_{t+1}$ is generated by the proposal distribution.
We use a Gaussian distribution which is centered at $\bz_t$ and has the fixed standard deviation of 0.1 as our proposal distribution.
A proposed sample is accepted with the probability of $\min \{1, \exp((E(\bar{\bz}_{t+1}) - E(\bz_{t}))/T)\}$, yielding $\bz_{t+1}=\bar{\bz}_{t+1}$. Otherwise, $\bz_{t+1}=\bz_{t}$.
We take the final state of a Markov chain as a generated sample.
As stated in the previous paragraph, the final state is rejected if $\bz\not\in\mathcal{T}_h$.
An accepted sample is mapped to $\mathcal{X}$ by $\bx=f(\bz)$.

\subsection{Variation Models}

\paragraph{Affine} ($D_\bz=5$) We use \texttt{torchvision.transforms.functional.affine} function to apply affine transform on an image. There are five degrees of freedom and their parameter bounds are selected as follows: angle (-45 $\sim$ 45), translation on x-axis (-10 $\sim$ 10), translation on y-axis (-10 $\sim$ 10), scale (0.9 $\sim$ 1.5), and shear (-30 $\sim$ 30). These parameters are set by hand. 

\paragraph{Color} ($D_\bz =4$) We use \texttt{adjust\_brightness}, \texttt{adjust\_contrast}, \texttt{adjust\_saturation}, \texttt{adjust\_hue} from \texttt{torchvision.transforms.functional}. Parameters and their bounds are given as follows: brightness (0.5 $\sim$ 1.5), contrast (0.5 $\sim$ 1.5), saturation (0 $\sim$ 2), and hue (-0.5 $\sim$ 0.5).

\begin{figure}[h]
    \centering
    \includegraphics[width=0.23\textwidth]{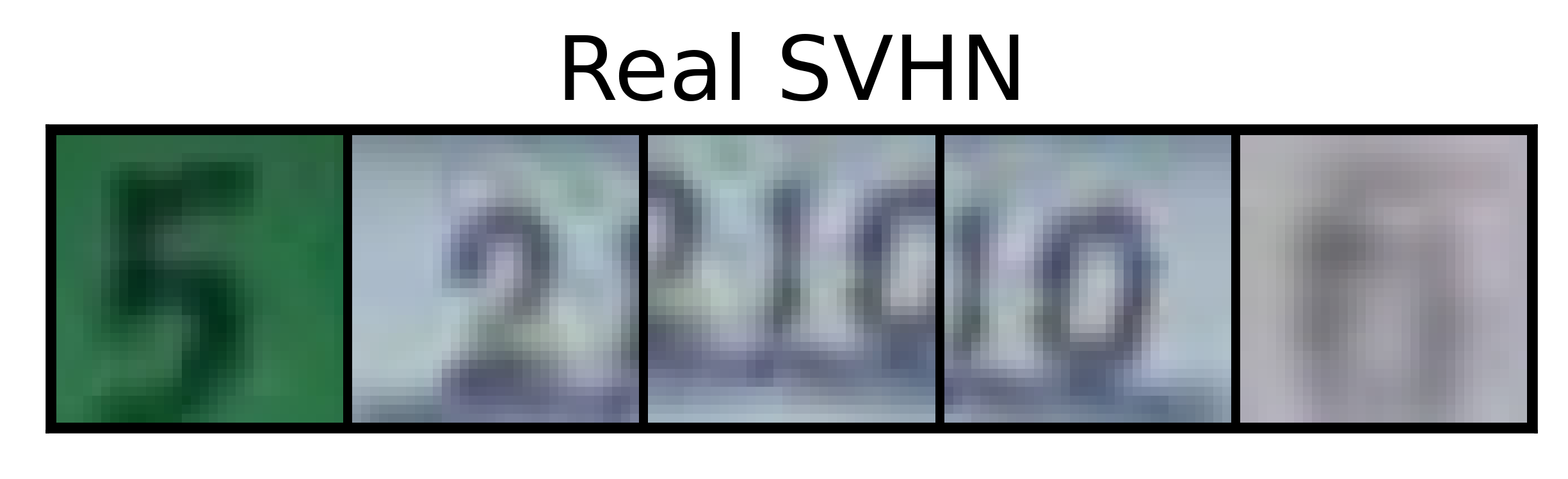}
    \includegraphics[width=0.23\textwidth]{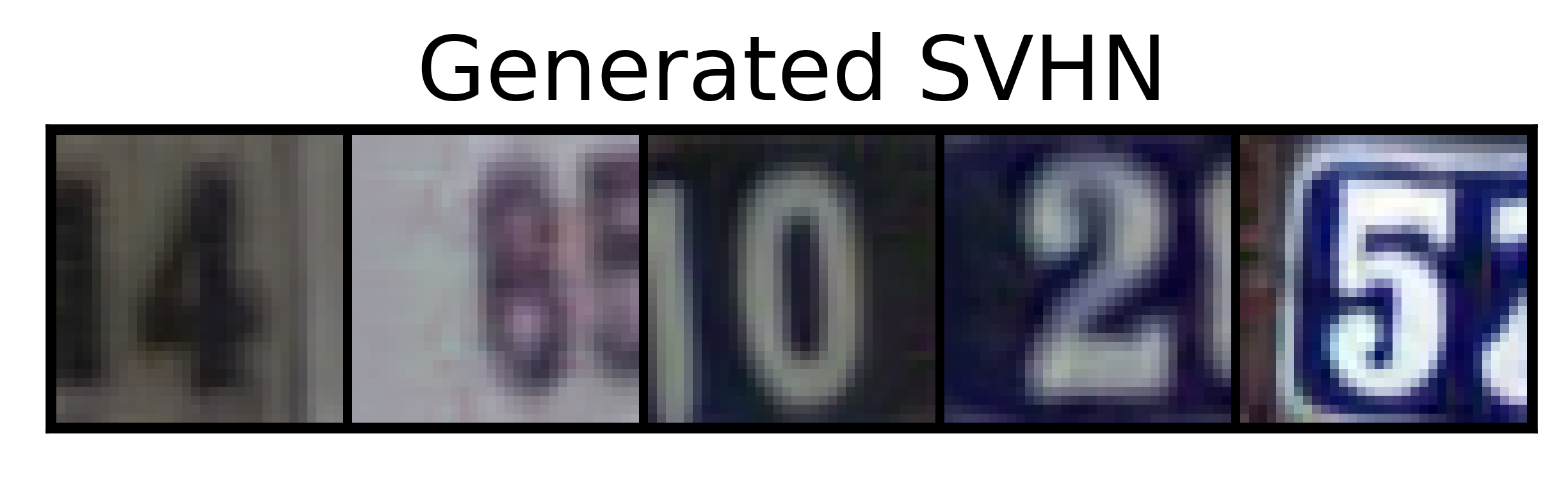}
    \includegraphics[width=0.23\textwidth]{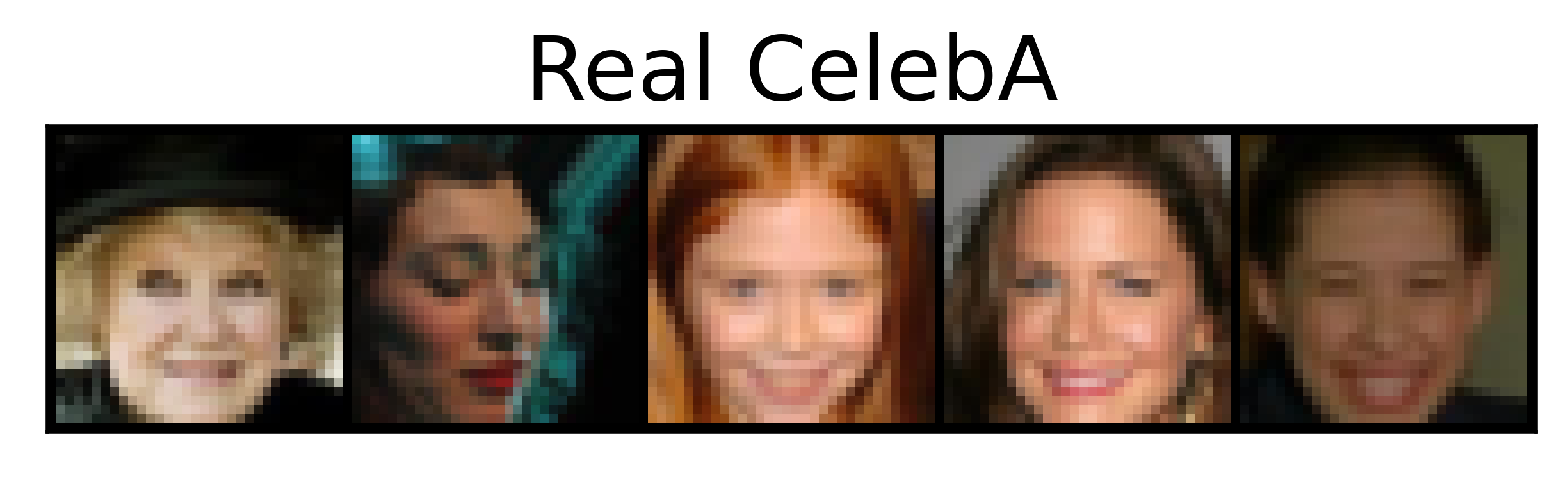}
    \includegraphics[width=0.23\textwidth]{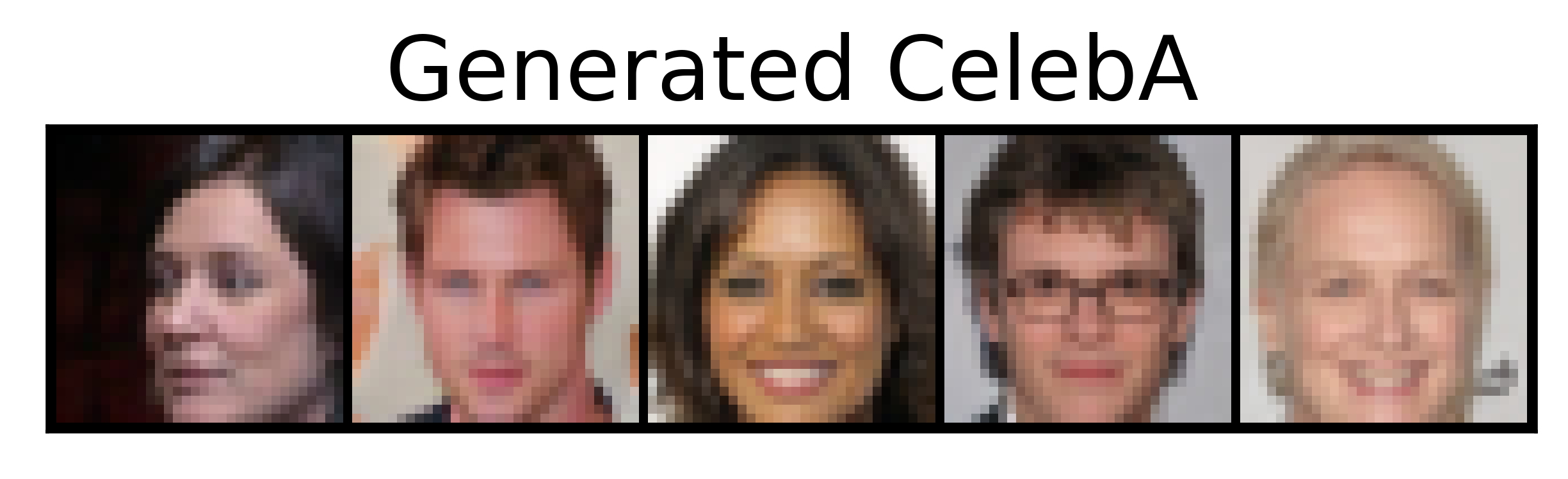}
    \caption{Comparison between real OOD images (SVHN and CelebA) and images generated from StyleGAN2 used in our experiments.}
    \label{fig:gan}
\end{figure}

\paragraph{StyleGAN2} ($D_\bz=64$) We train two StyleGAN2 networks, one for SVHN and one for CelebA. We use the default setting of PyTorch-StudioGAN repository\footnote{https://github.com/POSTECH-CVLab/PyTorch-StudioGAN} except that we change the latent dimensionality to 64. The training took 24 hours on two A6000 GPUs and went smoothly. The randomly generated samples from the trained GAN are shown in Figure \ref{fig:gan}. The samples are realistic and almost not distinguishable to the real samples.

\section{Implementation of OOD Detectors}

\subsection{Implementation of Weak and Strong Detectors}

In this section, we describe how OOD detectors are implemented in our experiments.

\paragraph{AE}
The architecture of the autoencoder is similar to what is used in \citep{Ghosh2020From} and is described in Table \ref{tb:ae} with a different dimensionality of the latent space $D_Z=128$.
AE is trained for 500 epochs on in-distribution training set using Adam optimizer with the learning rate $1\times10^{-4}$. A mini-batch contains 128 samples. A model with the smallest reconstruction error on validation split is selected.

\begin{table}[h]
    \centering
    \caption{Autoencoder architecture. Conv$_N$(M) indicates a 2D convolution operation with a $N\times N$ kernel and $M$ output channels. BN denotes batch normalization, and ReLU means the rectified linear unit activation. }
    \label{tb:ae}
    \begin{tabular}[t]{c}
        \toprule
         Encoder\\ 
         \midrule
         Conv$_4$(128)-BN-ReLU-\\
         Conv$_4$(256)-BN-ReLU-\\
         Conv$_4$(512)-BN-ReLU-\\
         Conv$_4$(1024)-BN-ReLU-\\
         FC(1024)-BN-ReLU-\\
         FC(128)-Tanh\\
         \bottomrule
    \end{tabular}
    \begin{tabular}[t]{c}
        \toprule
         Decoder\\ 
         \midrule
         ConvT$_8$(1024)-ReLU-\\
         ConvT$_4$(512)-ReLU-\\
         ConvT$_4$(256)-ReLU-\\
         ConvT$_1$(3)-Sigmoid\\
         \bottomrule
         
    \end{tabular}
\end{table}

\paragraph{PXCNN}
PXCNN is implemented based on an open-sourced code repository\footnote{\url{https://github.com/pclucas14/pixel-cnn-pp}} and re-trained on CIFAR-10. All the model parameters are set to the default values, i.e., \texttt{nr\_resnet=5, nr\_filters=80, nr\_logistic\_mix=10, resnet\_nonlinearity=`concat\_elu'}. An input image is linearly scaled into $[-1,1]$, and horizontal flipped with the probability of 50\%.
The model is trained for 200 epochs where a mini-batch contains 64 samples. 
We use Adam optimizer with the learning rate $1\times10^{-4}$ which is decayed by a multiplicative factor of 0.999995 every iteration.

\paragraph{Glow} 
Glow is implemented based on two repositories\footnote{\url{https://github.com/chaiyujin/glow-pytorch}}
\footnote{\url{https://github.com/chrischute/glow}} and traianed on CIFAR-10.
Our Glow model utilizes a multi-scale architecture with three levels of the latent representations are present, i.e., \texttt{L=3}. Each level contains 32 flow steps. We use 1$\times$1 invertible convolution and ActNorm with the scale of 1.0.
The model is trained for 500 epochs with the batch size of 64. Adam optimizer with the learning rate 1$\times10^{-4}$ is utilized. The gradient is clipped so that its norm is no larger than 0.1.


\paragraph{NAE} The pre-trained model is downloaded from the official repository\footnote{\url{https://github.com/swyoon/normalized-autoencoders}}.
We use Conv32Big version, which showed better performance on CIFAR-10 vs SVHN.

\paragraph{OE} Among the available pre-trained models in the official repository\footnote{\url{https://github.com/hendrycks/outlier-exposure}}, we select the best performing version, i.e., \texttt{cifar10\_allconv\_oe\_scratch\_epoch\_99.pt}. We normalize an input image with \texttt{means=(0.4914, 0.4822, 0.4465)} and \texttt{std=(0.2471, 0.2435, 0.2615)}, where each component corresponds to RGB, respectively.

\paragraph{CEDA, ACET, GOOD} The pre-trained models are provided by the official repository of GOOD\footnote{\url{https://gitlab.com/Bitterwolf/GOOD}}.
Among multiple versions of GOOD, we use GOOD80, as recommended in the original paper.

\paragraph{MD} We implement MD following the official repository\footnote{\url{https://github.com/pokaxpoka/deep_Mahalanobis_detector}}.
Based on the pre-trained ResNet-32 provided by the official repository, we train the weights of each layer where Mahalanobis distance is computed.
1,000 SVHN images are used to determine such weights. Note that using test OOD dataset during the training is usually not acceptable in a typical OOD detection setting.

\paragraph{SSD} We use the official training script to train SSD
\footnote{\url{https://github.com/inspire-group/SSD}}.

\paragraph{CSI} We download the unlabelled multi-class CIFAR-10 model from the official repository\footnote{\url{https://github.com/alinlab/CSI}}.

\subsection{Detectors Not Included in Experiments} \label{appendix:not-implemented}

We have considered other OOD detectors such as Likelihood Ratio \citep{ren2019}, Input Complexity \citep{Serra2020Input}, and Likelihood Regret \citep{xiao2020likelihood}.
However, we could not include them in our experiments for the following reasons. Meanwhile, we are planning to include ViT \cite{fort2021exploring} to our experiments in the open-source version.


\paragraph{Likelihood Ratio} While there is the official public repository written in TensorFlow\footnote{\url{https://github.com/google-research/google-research/tree/master/genomics_ood/images_ood}}, we have failed to reproduce the result in PyTorch.

\paragraph{Input Complexity} 
We confirm that the log-density estimates from generative models are correlated to the bits after compression, but fail to achieve AUC score higher than 0.9 in CIFAR-10 vs SVHN setting. 
For generative models, PXCNN and Glow are tested, and for image compression algorithms, PNG, JPEG2000, and FLIF are tried.
Our generative models are implemented as Section \ref{sec:app-impl}.
For PNG and JPEG2000, we use OpenCV\footnote{\url{https://opencv.org/}}, and for FLIF, we use imageio-flif\footnote{\url{https://codeberg.org/monilophyta/imageio-flif}}.
No combination of a generative model and a compression algorithm results in AUC higher than 0.9.


\paragraph{Likelihood Regret} 
The method is very slow, particularly because it does not support batch processing, i.e., only one sample can be processed at a time. We decide that the method is too slow to apply the adversarial distribution approach.

\paragraph{ViT} 
OOD detection using Vision Transformer \cite{fort2021exploring} will be added to the open-source version of the project. It requires extra resources to integrate into the current code base because its implementation is in JAX. However, we believe the integration is possible and will be completed very soon. Although ViT show impressive results in OOD detection, a recent investigation confirm that it is not adversarially robust against noise-like perturbations \cite{fort2022adversarial}. Therefore, we suspect that ViT could be vulnerable under the variation models proposed in this paper.

\end{document}